# Low-dimensional phonon transport effects in ultra-narrow, disordered graphene nanoribbons

Hossein Karamitaheri[1], Mahdi Pourfath[2,3], Hans Kosina[3], and Neophytos Neophytou[4,5]

[1]Department of Electrical Engineering, University of Kashan, Kashan 87317-51167, Iran
[2]School of Electrical and Computer Engineering, University of Tehran, Tehran 14395-515, Iran
[3]Institute for Microelectronics, Technical University of Vienna, Vienna, 1040, Austria
[4]School of Engineering, University of Warwick, Coventry, CV4 7AL, UK
[5]E-mail: N.Neophytou@warwick.ac.uk

## Abstract

We investigate the influence of low-dimensionality and disorder in phonon transport in ultra-narrow armchair graphene nanoribbons (GNRs) using non-equilibrium Green's function (NEGF) simulation techniques. We specifically focus on how different parts of the phonon spectrum are influenced by geometrical confinement and line edge roughness. Under ballistic conditions, phonons throughout the entire phonon energy spectrum contribute to thermal transport. With the introduction of line edge roughness, the phonon transmission is reduced, but in a manner which is significantly non-uniform throughout the spectrum. We identify four distinct behaviors within the phonon spectrum in the presence of disorder: i) the low-energy, low-wavevector acoustic branches have very long mean-free-paths and are affected the least by edge disorder, even in the case of ultra-narrow $W$=1nm wide GNRs; ii) energy regions that consist of a dense population of relatively 'flat' phonon modes (including the optical branches) are also not significantly affected, except in the case of the ultra-narrow $W$=1nm GNRs, in which case the transmission is reduced because of band mismatch along the phonon transport path; iii) 'quasi-acoustic' bands that lie within the intermediate region of the spectrum are strongly affected by disorder as this part of the spectrum is depleted of propagating phonon modes upon both confinement and disorder (resulting in sparse $E(q)$ phononic bandstructure), especially as the channel length increases; iv) the strongest reduction in phonon transmission is observed in energy regions that are composed of a small density of phonon modes, in which case roughness can introduce transport gaps that greatly increase with channel length. We show that in GNRs of widths as small as $W$=3nm, under moderate roughness, both the low-energy acoustic modes and dense regions of optical modes can retain semi-ballistic transport properties, even for channel lengths up to $L$=1μm. These modes tend to completely dominate thermal transport. Modes in the sparse regions of the spectrum, however, tend to fall into the localization regime, even for channel lengths as short as 10s of nanometers, despite their relatively high phonon group velocities.

# I. Introduction

The thermal properties of graphene nanostructures and low-dimensional channels in general is an important topic of nanoscience. Graphene nanoribbons (GNR) are one-dimensional structures that have attracted significant attention, both for fundamental research as well as for technological applications [1, 2, 3, 4, 5, 6, 7, 8, 9, 10, 11, 12, 13, 14]. Ultra-narrow GNRs have been shown to retain at some degree the remarkable thermal properties of graphene. However, the presence of edges can result in geometry dependent properties. The width, chirality, and the magnitude of edge disorder of the GNR, can strongly determine its electronic [15, 16, 17, 18] and heat transport properties [9, 10, 19, 20, 21].

Several works have shown that the transports properties of low-dimensional systems are significantly degraded by the introduction of scattering centers and localized states [9, 10, 22, 14, 23, 24, 25]. In the case of electronic transport, even a small degree of disorder can drastically reduce the electronic conductivity (especially in AGNRs rather than ZGNRs), even driving carriers into the localization regime and introduce 'effective' transmission bandgaps [15, 26, 27, 28]. Although the line edge roughness can have a similar effect on the thermal properties of GNRs, it has not yet been theoretically explored in depth. Carbon related materials such as graphene, nanotubes, and GNRs can have huge thermal conductivities in their pristine form, reaching values as high as of 3080-5150 W/m K at room temperature [29, 2]. Even a small degree of disorder, however, can drastically degrade this superior thermal conductivity.

Recent theoretical studies attempt to address the thermal properties of low-dimensional materials by employing a variety of models and techniques depending on the size of the channel, and the physical effects under consideration. Methods to investigate low-dimensional thermal transport vary from molecular dynamics [30, 31, 32, 25, 33, 34], the Boltzmann Transport Equation (BTE) for phonons using scattering rates based on the single mode relaxation time approximation (SMRTA) [35, 36, 37, 38, 39, 40, 41], the non-equilibrium Green's function (NEGF) method [14, 24, 20, 42, 43, 44, 45, 46], and the Landauer method [47, 48, 49, 50], but also even more simplified semi-analytical methods that employ the Casimir formula to extract boundary scattering rates by assigning a diffusive or specular nature to the boundaries [51, 52].

One of the reasons why the phonon transport properties of low-dimensional channels in general, and carbon based systems in particular, are recently receiving much attention is the fact that they show certain features that are distinct from bulk materials. Several experimental and theoretical works suggest that the thermal conductivity could deviate from Fourier's law [3, 12, 53]. It was observed that it grows monotonically with channel length before it saturates at large channel lengths, even lengths significantly larger than the average mean-free-path (MFP) [54, 8], an indication of a crossover from ballistic into diffusive transport regimes [55, 56]. A recent theoretical study showed that, in the case of pristine 1D channels, the thermal conductivity could even increase with confinement [57]. References [58, 59, 60], demonstrated that the thermal conductivity in 1D channels grows as a power-law function of the length and that roughness affects the value of the exponent of this dependence. In 2D graphene channels, on the other hand, the increase in thermal conductivity with channel length follows a logarithmic trend [8].

The major effect in limiting thermal conductivity in 1D channels, however, seems to be boundary scattering [24, 61, 9]. Two orders of magnitude reduction in thermal conductivity has been reported for several low-dimensional materials due to roughness compared to the pristine materials, which significantly improve their thermoelectric properties [61, 62, 14]. Specifically, with regard to GNRs, studies concluded that edge roughness in GNRs can indeed reduce the thermal conductivity by up to two orders of magnitude, depending on the assumptions made about the roughness amplitude and the autocorrelation length.

The phonon spectrum of ultra-narrow GNRs and 1D-dimensional channels in general, however, consists of various phonon modes and polarizations, which react differently in the presence of disorder (i.e. line edge roughness) and exhibit different mean-free-paths (MFPs) and localization lengths (LL). Despite the tremendous theoretical and experimental investigations of thermal conductivity in nanostructures, a study on how line edge disorder in 1D GNR channels affects phonon modes of different frequencies and wavevectors in the entire phonon spectrum is still lacking. What is also lacking is a study on what changes the phonon modes undergo in different parts of the spectrum under strong confinement, and how these changes affect thermal transport in the presence of line edge roughness. The few studies that attempt to address this issue for other 1D channels reach various and differing conclusions. A study on thermal transport in 1D Si nanowires, for example, indicated that line edge roughness scattering affects the thermal conductivity by introducing band mismatch in the optical region of the spectrum [24]. Different works attribute the reduction in thermal conductance to phonon localization and the appearance of non-propagating modes [23, 63, 64]. Specifically in the case of GNRs, it is indicated that the majority of eigenmodes are localized and do not contribute to thermal transport [9], whereas other studies suggest that heat transport is semi-ballistic [56].

In this work we theoretically investigate in detail the effect of line edge roughness and confinement in phonon transport in ultra-narrow armchair GNRs for the phonon modes of the entire energy spectrum independently. The basic conclusions of this study can be applied generically to all 1D systems. We employ the NEGF method [65, 66], which can take into account the exact geometry of the roughness without any underlying assumptions, while we describe the phonon spectrum atomistically using force constants. We show that in the presence of line edge roughness, all behaviors, i.e. band-mismatch, localization, ballisiticity, diffusion, appear, and all play a role in determining the overall thermal conductivity and its reduction under disorder. However, each effect applies to different parts of the spectrum, and each has different geometric dependence on the specific channel length and width. The paper is organized as follows: In Section II we describe the models and methods we employ to calculate the phonon spectrum and phonon transport. In Section III we present the results on the influence of line edge roughness on the phonon transmission in different parts of the phonon spectrum. More specifically, we show that the phonon spectrum can be split into four different parts which react differently to disorder: i) the dispersive quasi-ballistic low-wavevector acoustic modes, ii) relatively 'flat' but dense phonon mode regions, iii) 'quasi-acoustic' (or folded acoustic) dispersive regions, and iv) low-density phonon mode regions. Section IV discusses the effect of edge roughness and GNR width on the *thermal conductance*. We show that although phonon localization is observed for certain frequencies independent of the GNR width, the overall thermal conductance indicates localization behavior only in ultra-narrow channels of width $W$=1nm. Channels of widths greater than a few nanometers are overall diffusive, even at channel lengths of $L$>1μm. In Section V we extract the MFP and localization length for the GNR channels, and show how different parts of the spectrum become localized at different channel lengths. Section VI discusses the effects of disorder and confinement on the *thermal conductivity*, and finally Section VII summarizes and concludes the work.

## II. Methods

**II.a) Theory:** Under the harmonic approximation, the motion of atoms can be described by a dynamical matrix as:

$$D = [D_{3\times3}^{(ij)}] = \left[\frac{1}{\sqrt{M_i M_j}} \left\{ \begin{array}{ll} D_{ij} & i \neq j \\ -\sum_{l \neq i} D_{il} & i = j \end{array} \right. \right] \quad (1)$$

where $M_{i,j}$ is the atomic mass of the $i^{\text{th}}$, $j^{\text{th}}$ carbon atom (in this case all atoms have the same mass), and the dynamical matrix component between atoms '$i$' and '$j$' is given by:

$$D_{ij} = \begin{bmatrix} D_{xx}^{ij} & D_{xy}^{ij} & D_{xz}^{ij} \\ D_{yx}^{ij} & D_{yy}^{ij} & D_{yz}^{ij} \\ D_{zx}^{ij} & D_{zy}^{ij} & D_{zz}^{ij} \end{bmatrix} \quad (2)$$

where

$$D_{mn}^{ij} = \frac{\partial^2 U}{\partial r_m^i \partial r_n^j}, \quad i,j \in N_A \text{ and } m,n \in [x,y,z] \quad (3)$$

is the second derivative of the potential energy ($U$) after atoms '$i$' and '$j$' are slightly displaced along the *m*-axis and the *n*-axis ($\partial r_m^i$ and $\partial r_n^j$), respectively.

For setting up the dynamical matrix component between the $i^{\text{th}}$ and the $j^{\text{th}}$ carbon atoms, which are the $N^{\text{th}}$ nearest-neighbors of each other, we use the force constant method (FCM), involving interactions up to the fourth nearest-neighbor [67]. The force constant tensor is given by:

$$K_0^{(ij)} = \begin{bmatrix} \phi_r^{(N)} & 0 & 0 \\ 0 & \phi_{ti}^{(N)} & 0 \\ 0 & 0 & \phi_{to}^{(N)} \end{bmatrix} \qquad (4)$$

where $\phi_r^{(N)}$, $\phi_{ti}^{(N)}$, and $\phi_{to}^{(N)}$ are the radial, the in-plane transverse, and the out-of-plane transverse components respectively. The force constant fitting parameters are taken from Ref. [68] and are shown to accurately reproduce the phonon dispersion of graphene. The 3x3 components of the dynamical matrix are then computed as:

$$D_{ij} = U_m^{-1} K_0^{(ij)} U_m \qquad (5)$$

where $U_m$ is a unitary rotation matrix defined as:

$$U_m = \begin{bmatrix} \cos\theta_{ij} & \sin\theta_{ij} & 0 \\ -\sin\theta_{ij} & \cos\theta_{ij} & 0 \\ 0 & 0 & 1 \end{bmatrix} \qquad (6)$$

Assuming the graphene sheet is located in the *x-y* plane, $\theta_{ij}$ represents the angle between the *x*-axes and the bond between the $i^{th}$ and $j^{th}$ carbon atom.

The phonon dispersion can be computed by solving the following eigenvalue problem:

$$\left[ D + \sum_l D_l \exp(i\vec{q}.\Delta\vec{R}) \right] \psi(\vec{q}) = \omega^2(\vec{q}) \psi(\vec{q}) \qquad (7)$$

where $D_l$ is the dynamical matrix representing the interaction between the unit cell and its neighboring unit cells separated by $\Delta\vec{R}$, and $\psi(\vec{q})$ is the phonon mode eigenfunction at wavevector $\vec{q}$.

The FCM is coupled to NEGF for the calculation of the coherent phonon transmission function in the GNR. The NEGF method is appropriate for studies of phonon transport in geometries with disorder because the exact geometry is included in the construction of the dynamical matrix. Employing an atomistic approach that considers the discrete nature of the line edge roughness and accurately models its impact on phonon modes is essential for the analysis of thermal properties of narrow GNRs (with $W<20$nm). The method considers the wave nature of phonons, rather than their particle description, and all interference and localization effects, which could be important in low-dimensional channels, are captured. In addition, it is most appropriate for the purposes of this study, which investigates the influence of line edge roughness for phonons of different frequencies of the spectrum, as NEGF computes the energy resolved phonon transmission function. The system geometry consists of two semi-infinite contacts made of pristine GNRs, surrounding the channel in which we introduce line edge roughness. The Green's function is given by:

$$G(E) = \left[ E^2 I - D - \Sigma_1 - \Sigma_2 \right]^{-1} \quad (8)$$

where $D$ is device dynamical matrix and $E = \hbar\omega$ is the phonon energy. The contact self-energy matrices $\Sigma_{1,2}$ are calculated using the Sancho-Rubio iterative scheme. The transmission probability through the channel can be obtained using the relation:

$$T_{ph}(\omega) = Trace\left[ \Gamma_1 G \Gamma_2 G^+ \right] \quad (9)$$

where $G_1$ and $G_2$ are the broadening functions of the two contacts defined as $\Gamma_{1,2} = i\left[ \Sigma_{1,2} - \Sigma_{1,2}^+ \right]$. The thermal conductance can then be calculated in the framework of the Landauer formalism as:

$$K_l = \frac{1}{2\pi\hbar} \int_0^\infty T_{ph}(\omega) \hbar\omega \left( \frac{\partial n(\omega)}{\partial T} \right) d(\hbar\omega) \quad (10)$$

where $n(\omega)$ is the Bose-Einstein distribution and $T$ is the temperature. In this work we consider room temperature $T=300$K. At room temperature and under ballistic conditions the function inside the integral spans the entire energy spectrum [69, 57], which allows phonons of all energies to contribute to the thermal conductance.

**II.b) Dispersion features:** Figures 1a and 1b show typical dispersion relations for GNR channels of widths $W=5$nm and $W=1$nm, respectively. The $W=1$nm case, as we show below, resembles purely 1D features, whereas at a width of $W=5$nm the dispersion diverts towards 2D (although the dispersions in both cases are 1D). These two sizes are computationally manageable, and comparison between their transport properties allows comparison between 1D and less confined, 'towards 2D', phonon transport. The colormap in Fig. 1 shows the contribution of each phonon state to the ballistic thermal conductance at room temperature. To analyze the observed features of the GNR phonon dispersions, let us first consider the graphene phonon dispersion. In graphene, there are 6 phonon modes, 3 acoustic and 3 optical modes [68]. The highest frequency acoustic mode is the longitudinal acoustic (LA) mode, the next one is the in-plane transverse acoustic mode (TA) and lowest frequency mode is the out-of plane acoustic mode (ZA). The latter is recently shown to make the largest contribution to the thermal conductivity of graphene [4, 5, 70, 71, 72]. The highest frequency optical mode is the longitudinal optical (LO), followed by the in-plane transverse optical (TO), and the lowest is the out-of-plane optical (ZO) [42, 72]. The LA mode of the GNRs shown in Fig. 1 is the corresponding LA mode of graphene with group velocity $v_s$=19.8 km/s. The LA and TA modes are linear at low frequencies, and extend up to $E$~0.16eV and $E$~0.14eV, respectively. The ZA mode is quadratic for low frequencies and extends up to $E$~0.07eV. At the higher part of their energy region, the acoustic modes become relatively 'flat'. The ZO modes extend from $E$~0.7eV-0.11eV, whereas the LO and TO modes are located at higher energies, from $E$~0.16eV-0.2eV. The relatively 'flat' mode regions around energies $E$~0.07eV-0.11eV consist of ZO modes, in addition to the dispersive LA and TA modes [42]. The less dispersive modes located from $E$~0.11eV-0.16eV are the 'flat' parts of the LA and TA modes.

## III. Effects of confinement and line edge roughness scattering

**III. a) Confinement effects on bandstructure:** Three main observations on the phonon bandstructure can be made as the width is reduced, i.e. between Fig. 1a and Fig. 1b:

i) The optical and 'quasi-acoustic' modes (which are nothing else but folded acoustic branches of the host material [73]) show strong confinement dependence [74]. The number of modes depends on the number of atoms within the unit cell. As the width is reduced from $W$=5nm (Fig. 1a) to $W$=1nm (Fig. 1b), the number of modes in these regions is also reduced.

ii) The number of acoustic modes remains intact, and they carry a much larger portion of the heat (as indicated by their red coloring in Fig. 1a and 1b).

iii) Small bandgaps appear in some regions in the bandstructure, especially in regions around the interface between the 'flat' optical modes and the more dispersive 'quasi-acoustic' modes (primarily around $\hbar\omega \sim 0.16$ eV, and secondly around $\hbar\omega \sim 0.11$ eV, and $\hbar\omega \sim 0.07$ eV). In addition, large regions in the phononic $(\hbar\omega, q)$ space, especially in the 'quasi-acoustic' band regions, become 'empty' of modes (sparse), where for rather extensive energy and momentum intervals no phonon states exist.

**III. b) Effect of roughness on phonon transmission:** We then investigate phonon transport in these low-dimensional GNRs in the presence of disorder. At such small ribbon widths with rough edges, the edge-phonon scattering is the dominant scattering mechanism [25]. For this, we simulate rough GNR channels of width $W$=5nm (relatively wide) down to $W$=1nm (purely 1D), and examine the phonon transmission across the phonon energy spectrum as the length of the GNR increases (i.e. as the effective disorder increases). We construct the line edge roughness (LER) geometry by adding/subtracting carbon atoms from the edges of the pristine GNR according to the exponential autocorrelation function:

$$R(x) = \Delta W^2 \exp\left(-\frac{|x|}{\Delta L}\right) \qquad (11)$$

where $\Delta W$ is the root mean square of the roughness amplitude and $\Delta L$ is the roughness correlation length [26]. The Fourier transform of the autocorrelation is the power spectrum of the roughness. The real space representation of the LER is achieved by adding a random phase to the power spectrum followed by an inverse Fourier transform [26, 75]. We use $\Delta W = 0.1 nm$ and $\Delta L = 2 nm$. We keep this roughness description constant in all cases. Therefore, the 'effective' disorder in the channels we simulate increases as: i) the channel length is increased, or ii) the channel width is reduced. In the results that follow, for every channel GNR of different length/width, we average over 50 realizations of different channels.

Figure 2 shows the transmission function of the phonon spectrum as a function of energy for the GNR with width $W$=5nm (Fig. 2a), and for the ultra-narrow GNR of width $W$=1nm (Fig. 2b). The figure shows transmissions of channels with rough edges and various lengths. The dashed-black lines indicate the ballistic transmission of the GNRs with perfect edges. The transmission of GNRs with length $L$=5nm (blue line), $L$=40nm (red line), $L$=100nm (green line) and $L$=500nm (black-solid line) are plotted.

The transmission is significant in the entire energy spectrum and thus the whole spectrum contributes to thermal conductance for both the wide and narrow GNRs [26]. Of particular note is the sharp transmission peak in the high energy optical modes in the case of the wide GNR in Fig. 2a, which originates from their large number, rather than their group velocity, which is low. Line edge roughness reduces the transmission function significantly, and in particular around energies $E$=0.06eV-0.07eV, $E$=0.11eV-0.14eV, and $E$=0.16eV-0.17eV. This group of energy regions, for which the transmission is strongly reduced, are regions of low density (but also dispersive) modes. In particular, the latter energy region is the one around the boundary between 'flat' and 'dispersive' modes, exactly above the energy at which the LA mode ends, and is a region with particularly low mode density. A surviving contribution to the transmission is evident around energies $E$=0-0.05eV (acoustic phonons), $E$=0.08eV-0.11eV (a mixture of LA, TA, and ZO modes), and $E$=0.17eV-0.2eV (optical phonons), even for the longer length GNRs. It is evident from this that the low group velocity optical modes contribute significantly to transmission due to their large density, even at the presence of roughness.

The corresponding transmission functions for the narrower GNR with width $W$=1nm shown in Fig. 2b, undergo much stronger reductions with line edge roughness compared to the wider GNRs of the same length. Since we keep the roughness amplitude the same in all cases, reducing the width essentially increases the effective disorder. The reduction is much stronger in the entire energy spectrum, in particular around the low density mode energy regions ($E$=0.06eV-0.07eV, $E$=0.11eV-0.14eV, and $E$=0.16eV-0.17eV as mentioned above), where the transmission is diminished. What dominates thermal conductance in the ultra-narrow GNR case, especially when the length of the channel is increased above $L>40$nm, are the low-energy, low-wavevector acoustic modes (black solid line in Fig. 2b). This is clearly indicated in the inset of Fig. 2b, which shows in logarithmic scale the transmission of the ballistic GNR channel and the transmission of the rough edge GNR channel with $L=1\mu$m and $W=1$nm. Clearly, only the transmission in the low-energy region survives.

**III. c) Effects of roughness on different phonon modes:** To illustrate the distinctly different behavior of the various phonon modes in the presence of line edge roughness, Fig. 3 shows the transmission at certain phonon frequencies as a function of the channel length $L$. Figures 3a, 3b, and 3c show results for the $W=5$nm, $W=3$nm, and $W=1$nm GNRs, respectively. We concentrate on four different phonon categories, and pick a specific phonon energy within the energy region of these categories. These are: i) acoustic phonons ($E=0.01$eV, blue lines), ii) optical, 'flat' dispersion phonons ($E=0.19$eV – red solid lines, and $E=0.09$eV – red-dashed lines), iii) 'quasi-acoustic', dispersive phonon modes ($E=0.13$eV, black lines), and iv) regions of very low mode densities, in which confinement can even result in narrow bandgaps ($E=0.16$eV, green lines). For all energy cases, and for all GNR widths, the transmission drops with increasing channel length and reducing width. The drop, however, differs significantly for each different phonon energy case. The drop in the transmission of the acoustic modes (blue lines) is relatively weak, and can be understood from the fact that they are composed of LA modes with long wavevectors [10, 11]. These modes are very weakly affected by defects, and this is the case for both wider and ultra-narrow GNRs. For example, Scuracchio *et* al. have also indicated that these modes are only weakly affected by atomic vacancies [76], and Huang *et* al. reached very similar conclusions in the presence of dislocation defects in GNRs [77]. The optical modes (red-solid and red-dashed lines), have a much stronger dependence on the GNR width. For the wider channel (Fig. 3a), their transmission is even larger compared to the acoustic modes independent of channel length. As the width is reduced, their transmission drops with increasing length, especially in the case of the ultra-narrow $W=1$nm channel. In the case of the 'quasi-acoustic' modes (black lines), a large drop in the transmission is observed as the channel length increases. Even stronger is the drop in the transmission of the very low density mode regions (green lines). In the following sections, we provide explanations regarding this behavior.

**III. d) Ballistic, diffusive, localized modes:** In recent experiments in graphene and carbon nanotubes it was shown that thermal transport could deviate from Fourier's law and exhibit semi-ballistic behavior [6, 8]. Since each phonon mode responds differently to disorder, it is essential to investigate the regions of operation of the different modes, and identify the ones that contribute to the semi-ballistic behavior. Figures 3d, 3e and 3f, show the product of the transmission times the length of the channel ($T \times L$) versus channel length $L$ for the same channels and phonon modes as in Fig. 3a, 3b, and 3c, respectively. In the case of ballistic transport, the $T \times L$ product increases linearly. In the case of diffusive transport it remains constant. In the case of sub-diffusive transport the product reduces with length [78, 79, 80], and for localized transport, the product drops exponentially. From Fig. 3d and 3e, it can be observed that for the wider GNR channels, the acoustic modes (blue lines) are semi-ballistic, even for channel widths $W$=3nm and lengths up to $L$=1μm. For the ultra-narrow $W$=1nm GNRs (Fig. 3f), the acoustic modes reach the diffusive regime at around lengths of $L$~200nm, and get into the localized regime for lengths larger than $L$~700nm. Interestingly, a similar trend is observed for the optical modes (red lines) as well. For GNR widths $W$=5nm (Fig. 3d) and $W$=3nm (Fig. 3e), they indicate a semi-ballistic behavior even up to channel lengths of hundreds of nanometers. In the $W$=1nm case, though, the optical modes reach the localization regime at lengths well below $L$~100nm. The behavior of the 'quasi-acoustic' modes (black lines), on the other hand, is very different. These modes enter the diffusive regime at much shorter channel lengths compared to the acoustic and the optical modes. They even enter the localization regime after $L$~300nm for the $W$=5nm GNRs, after $L$~100nm for the $W$=3nm GNRs, and just after $L$~10nm for the $W$=1nm GNRs. This is quite intriguing since these are dispersive modes with much higher group velocities than the optical modes. The strongest reduction in transmission, however, is observed for the energy regions of low mode density (green lines). For these modes, the transmission is completely diminished after channel lengths of $L$~100nm in the case of the wider channels, and after $L$~10nm in the case of the ultra-narrow channel.

To clarify the diffusion-localization crossover, and demonstrate that the modes at energies $E\sim0.13$eV and $E\sim0.16$eV are actually into the localization regime, we plot the transmission fluctuations and histograms extracted from a large number of simulated samples. The phonon-transmission fluctuation is defined by a standard deviation:

$$\Delta T = \sqrt{\langle T^2 \rangle - \langle T \rangle^2}, \qquad (12)$$

which differs in the diffusive and localization regimes. In the case of diffusive transport the transmission histograms are described by a Gaussian distribution function [15] and the standard deviation is independent of the phonon energy [81]. In other words, the conductance fluctuation in the diffusive regime is universal, and the universal value is $\Delta T=0.365$ [15, 81]. In the ballistic and localization regimes, on the other hand, the so-called universal phonon-transmission fluctuation is not realized, and the standard deviations deviate from $\Delta T=0.365$. Specifically in the localization regime, the transmission histograms are described by a log-normal distribution function [81]. In the ballistic regime, the histograms as we show below are very narrow, centered just below the pristine channel ballistic transmission value.

Figure 4a shows the transmission standard deviation for GNRs with a width of $W=3$ nm and lengths of $L=100$ nm and $L=250$ nm. The value of universal phonon transmission fluctuation ($\Delta T=0.365$) is indicated by the horizontal dotted line [81]. To construct this figure, data from 8000 simulations for channels $L=100$nm and 1100 simulations for channels with $L=250$nm were used. In the case of low energy acoustic phonons, the mean-free-path is relatively large and their transport is ballistic (see the curve for $E=0.01$ eV in Fig. 3e), which results in small transmission fluctuations. As the energy increases to values around from 0.01eV to 0.05eV, transport becomes diffusive (the fluctuations are around the universal diffusive value shown by the dotted line). For energies around $E\sim0.07$eV and around $E\sim0.13$eV, transport enters the weak localization regime, and the fluctuations drop. The lowest amount of fluctuations is observed around energies of $E\sim0.16$eV, due to the fact that transport enters the strong localized regime (note that strong localization and ballistic regimes have both low fluctuations for different reasons). Finally, very close to diffusive transport is realized for the large energy optical phonons around $E$~0.19eV, where the deviation of the transmission fluctuations approaches the universal value again.

Figures 4b, 4c and 4d show the histograms of the transmission for various energies in the phonon spectrum. Figure 4b shows the histograms at channel lengths $L$=250nm for energy $E$=0.002eV which illustrates ballistic behavior, and energies $E$=0.02eV, and $E$=0.04eV, which illustrate diffusive behavior. The simulation data is indicated with dots, whereas the blue lines are Gaussian distributions plotted using the average and standard deviation of the simulation results. The standard deviation in the two cases is $\Delta T$=0.332 and $\Delta T$=0.339, values very close to the universal fluctuation value, $\Delta T$=0.365. Note the sharp distribution in the case of ballistic transport, indicating that disorder does not affect the transport of the very low energy acoustic modes. The phonon mode at $E$=0.16eV, on the other hand, is fully localized as indicated above. Figure 4c shows the transmission histograms at $E$=0.16eV in logarithmic scale for channel lengths $L$=100nm. The distribution function is clearly log-normal, indicating that the transport at that energy is completely localized. Finally, Fig. 4d shows the histogram of transmission at $E$=0.19eV for $L$=100nm channels, which follows a Gaussian distribution with a standard deviation of 0.31, again indicating a diffusive regime. We note that a very similar behavior is observed for phonons around energy $E$=0.09eV as well.

As discussed above in Fig. 3e, at GNR channel lengths $L$=100nm, for phonons at energies $E$=0.07eV, 0.09eV, 0.13eV, and 0.19eV the transport is nearly diffusive or weakly localized, therefore the fluctuation is close to the universal value, as also indicated by the blue line in Fig. 4a. As the length increases to $L$=250 nm, more phonon modes gradually enter in the localization regime (especially the modes around $E$=0.07eV and $E$=0.13eV) and the fluctuations deviate from the universal one. The conductance fluctuation histograms (not shown) for these two energies begin to resemble log-normal distributions at channel lengths of $L$~100nm. This is an indication that at this channel length these modes are at the beginning of the localization regime, as also shown in Fig. 3e. For channel lengths $L$=250nm and $L$=500nm, the distributions are very close to log-normal. For channels with $L$=250nm the standard deviations are $\Delta T$=0.199 and $\Delta T$=0.285 for the energies $E$=0.07 and $E$=0.13, respectively. As the channel length increases to $L$=500nm, the respective standard deviations decrease to $\Delta T$=0.056 and $\Delta T$=0.177, and indication of stronger localization. The lower deviation for the phonon at $E$=0.07eV is an indication of stronger localization at this energy compared to the phonons at $E$=0.13eV.

It should be noted that the localization appears only in the phase-coherent transport regime [82]. In the presence of phase-breaking phenomena, however, the localized states are removed and transport returns to the diffusive regime [83, 84]. For phonons, dephasing can be primarily due to phonon-phonon, and secondly due to electron-phonon interactions, neither of which we do not consider in this study. Localization will appear only if the phonon coherence length becomes longer than the localization length. Several works in the literature report the phonon-phonon scattering mean free path in graphene at room temperature to be in the range from a few to several hundreds of nanometers [29, 54, 56, 85]. We discuss the implications of this in detail for the structures we consider in Section V below.

**III. e) Transmission features in width modulated GNRs:** In Figures 5 and 6 we provide explanations for the behavior of the transmission in the different phonon energy regions with channel length and width. We base our analysis on two effects that explain the behavior of the modes: i) the change in the phonon bandstructure at specific energies under the influence of roughness, and ii) the corresponding change under the influence of geometrical confinement. We demonstrate that increasing effective roughness has a similar effect as increasing confinement. For example, regions in the phonon spectrum that become sparse of modes due to confinement, tend to more easily form 'effective' bandgaps in the presence of roughness as well, driving the transmission into localization. Figure 5 discusses the effect of roughness on specific energy regions of the bandstructure, whereas Fig. 6 the effect of roughness specifically on the sparse mode regions.

In Figure 5 we consider the $W$=1nm GNR and the following situation: We simulate the phonon modes and transmission for the ultra-narrow GNR of width $W$=1.1nm, a GNR of width $W$=0.74nm, and a GNR whose width is periodically modulated along its length (rather than randomly as in the case of rough channels), as shown in Fig. 5f (lower blue sub-figure). In this case, we can isolate the influence of roughness on the bandstructure. The left panels of the sub-figures of Fig. 5 show the phonon bandstructure of the three channels in the vicinity of the energies of interest. The bandstructure for the wide channel is shown in red, for the narrow channel in green, and for the width modulated channel in blue. The corresponding right panels show the transmission of the three channels. Figures 5a-e show, respectively, results for energies around $E$=0.001eV (low-frequency acoustic modes), $E$=0.09eV and $E$=0.19eV (optical modes), $E$=0.16eV (low density region modes), and $E$=0.13eV ('quasi-acoustic' modes).

**Acoustic modes:** In the case of the low-frequency acoustic modes in Fig. 5a, the transmission of the modulated channel is dominated by the transmission of the narrow region. In a small energy range a band mismatch is observed around the edge of the Brillouin zone, and the transmission is further reduced. In general, however, the reduction in transmission is relatively weak, which explains the fact that these modes behave semi-ballistically, especially as the energy and wavevector approach zero.

**Optical modes:** In the case of optical, 'flat' dispersion modes around energies $E$~0.09eV and $E$~0.19eV, it is evident from Fig. 5b and Fig. 5c that the reduction in the transmission due to width modulation (or roughness), originates from a band mismatch between the narrow and wider GNRs. The transmission of the width-modulated GNR is actually lower compared to the transmissions of both the wide and the narrow GNRs. For this $W$=1nm GNR, the density of optical modes is rather low, and the mismatch that is created under width modulation along the length of the channel can be significant, which degrades the transmission.

**Low-density mode regions:** Figure 5d shows the width-modulated results for the low-density mode energy regions at energies around $E$~0.16eV. As in the case of optical modes, a strong mismatch can be observed between the bands of the width-modulated GNR and the bands of both the wide and narrow GNRs. The mismatch, however, is much larger, at a degree where energy bandgaps are formed in the transmission function (Fig. 5d, right panel). Note that small bandgaps are also formed even in the uniform channels under strong confinement around this energy, which further increases the band mismatch in the presence of line edge roughness. The combination of bandgap formation and band mismatch justifies the drastic transmission drop for this particular energy region as the channel length increases (see for example Fig. 3, green lines).

**'Quasi'-acoustic modes:** Moving along to the case of the 'quasi-acoustic' modes of energy $E\sim0.13$eV shown in Fig. 5e, it is evident that the bands of the width-modulated GNR can look quite different compared to the bands of the wide or narrow GNRs. Some mode mismatch can be observed, which reduces the transmission even down to zero in certain parts of the spectrum. This, however, only partially explains why the drop with channel length shown in Fig. 3 (black lines) is so strong, i.e. it is much stronger compared to the drop in the optical modes at energy $E\sim0.09$eV or $E\sim0.19$eV.

The reason why the 'quasi-acoustic' modes behave so drastically different compared to the optical modes, can be explained by the looking at their behavior under confinement. Figure 1 shows that under confinement, the number of modes in these energy regions ($E\sim0.07$, and $E\sim0.13$) is reduced significantly, making these regions to look almost 'empty' of modes. In the presence of line edge roughness in a real geometry, the sparsity of the modes makes these particular energy regions more susceptible to the formation of 'effective' bandgaps by increasing the band mismatch. Such an event is not the case for the optical modes for the geometries we examine. The 'effective' transmission bandgap formation is demonstrated in the transmission functions shown in logarithmic scale in Fig. 6. Figure 6a shows the logarithmic transmission of the $W=1$nm GNR under ballistic (pristine channel) conditions (black line) and under line edge roughness when the channel length is $L=40$nm (red line). It is evident that for energies around $E\sim0.07$eV and $E\sim0.13$eV large 'effective' bandgaps form as indicated by the arrows, which become wider as the channel length increases even further (not shown). Figure 6b shows the same transmissions for the $W=5$nm GNR, but in this case we also plot the transmission for the GNR with $L=500$nm as well (green line). For short channels, the transmission is not significantly disturbed, but for the longer channels, bandgaps similar to the ones of the $W=1$nm GNR of Fig. 6a form around $E\sim0.07$eV and $E\sim0.13$eV, as also indicated by the arrows. Notice the even larger bandgap formation at energies $E\sim0.16$eV. This clearly indicates that the energy regions which become sparse of modes under confinement, are very susceptible to roughness in less confined geometries as well, which suggests that the influence of confinement has similar features in the transmission as the effect of roughness.

The behavior described above should hold for any sparse mode energy regions. Note, for example, that gaps do not form in the regions of the 'flat' optical modes, and the transmission does not degrade as much. Under strong confinement, however, the 'flat' optical mode regions become sparser, and in extreme cases begin to 'look' like the low-density regions as well. Under these conditions, they could also be subject to the effect we describe above. In this context, the thermal conductivity is a function of the width-dependent phonon spectra [25], for which line edge roughness could either further increase the band mismatch, or form 'effective' transport bandgaps.

We mention here that as in the case of electronic transport, the chirality (or 'aromaticity' [86]) of GNRs, i.e. armchair (AGNRs), or zig-zag (ZGNRs) can provide anisotropy in phonon transport behavior (although smaller compared to electronic transport anisotropy). In Ref. [87], for example, using the phonon Boltzmann transport equation, it was shown that the amount of anisotropy between AGNR and ZGNR ribbons can be significant, and increases as the ribbon width decreases and as the roughness amplitude increases. In the Appendix we show how the bands and the transmission of the ZGNR change under confinement and roughness, and compare this behavior to the corresponding AGNRs, indicating very similar qualitative behavior. An important message we convey in this work, however, is the fact that just by looking at how the phonon bandstructure behaves under confinement, and at its low-dimensional dispersion features, one can provide an indication of how the modes will behave under edge roughness. We do not focus specifically on the details of the GNR dispersion itself, but we rather provide general low-dimensional phonon transport features. Qualitatively, the behavior we describe should hold for other low-dimensional materials, but could also be relevant to graphene ribbon phonon dispersions extracted through DFT calculations (using LDA, GGA, or GW which can produce slightly different dispersions with respect to each other), and might also produce slightly different dispersions compared to the ones obtained using the force constant method we employ here. Indeed, several works have investigated the phonon dispersions and phonon localization in graphene nanoribbons using DFT calculations [88, 89, 90, 91], with mainly similar observations. In our previous works we have shown that the force-constant-method (as a semi-empirical method with fitting parameters) can correctly regenerate the bandstructure of graphene, obtained from first-principle calculations [67]. Furthermore, we have also shown that by employing this approach for a relative roughness between ~0.5% and ~5% of ribbon's width, a very good agreement with the experimental data for GNRs with widths up to ~15nm can be achieved [26]. Thus we trust that the dispersions we employ are accurate enough compared to more sophisticated DFT calculations. In any case, to properly account for transport properties, we treat roughness atomistically, which is essential to study transport in narrow ribbons. We consider channels with lengths of about 1 μm that result in more than 10,000 atoms, which would make the use of DFT (combined with Green's function transport calculations) almost computationally impossible, whereas the force constant method provides a feasible way to study transport in relatively long, rough channels.

## IV. Thermal conductance

**IV. a) Thermal conductance:** We next consider the thermal conductance of the GNRs at $T$=300K in the presence of line edge roughness. We consider channels of different widths and lengths as shown in Fig. 7a. The thermal conductance drops as the channel lengths increases, and the reduction rate, if compared to Fig. 3a-c, follows the reduction in the transmission of the dominant modes. For the wider GNRs, the reduction rate is smaller, as the transmission of the dominant acoustic and optical bands is affected only slightly. As the width is reduced down to the ultra-narrow $W$=1nm, the thermal conductance drops faster with channel length (blue-dotted line).

Interestingly, by plotting the product of thermal conductance times channel length $K \times L$ in Fig. 7b, we show that only the wider channel with $W$=5nm operates in the quasi-ballistic regime ($K \times L$ continues to increase even up to channel lengths of $L$=750nm). The channels with widths $W$=4nm, 3nm and 2nm operate in the diffusive regime for channel lengths beyond $L$=500nm ($K \times L$ saturates to a constant value). The ultra-narrow $W$=1nm channel, on the other hand, for channel lengths $L$>300nm enters the localization regime ($K \times L$ decreases – see inset of Fig. 7b). In either channel case, modes exist that are ballistic, diffusive, or localized as discussed above. The overall behavior at larger channel lengths, however, is dominated by the behavior of the acoustic modes (the wider GNRs have a strong contribution from the optical modes as well).

**IV. b) Cumulative thermal conductance:** The dominance of the acoustic modes is clearly illustrated in Fig. 8a-c, which shows the cumulative thermal conductance at room temperature as a function of energy for the GNRs of widths $W$=5nm, $W$=3nm and $W$=1nm, respectively. Results for GNRs of lengths $L$=5nm (blue lines), 40nm (red lines), 100nm (green lines), and 500nm (black lines) are shown. By the dashed-dot lines we show the cumulative ballistic thermal conductance. In the ballistic case, independent of the GNR width, the entire spectrum contributes to thermal conductance, with the low energy acoustic modes contributing ~50%, and the high energy optical modes ~10%, whereas the rest ~40% is contributed from phonons in the intermediate energy region. For the roughened wider GNR with $W$=5nm (Fig. 8a), this behavior is also independent of channel length, and retained until at least $L$=500nm. As the width of the GNR is reduced, i.e. in the $W$=3nm GNR case shown in Fig. 8b, the situation is similar, except that at larger channel lengths, the contribution of the low energy phonons increases. The higher energy modes get into sub-diffusion and/or localization regimes and contribute less. This results in ~80% of the heat to be carried by phonons with energies below $E$=0.02eV. For even narrower GNRs, as the ultra-narrow $W$=1nm GNR shown in Fig. 8c, the distribution shifts towards the low energy acoustic modes at much shorted channel lengths, even as short as $L$=5nm (blue line). In the limit of very long and very narrow channels, i.e. approaching purely 1D, all heat is carried by the very low energy acoustic modes, whereas all higher energy modes are driven into the localization regime [6, 57].

## V. Mean-free-path and localization length

To identify the dependence of the transmission function on the channel length for the different operating regimes, we need to relate it to the mean-free-path (MFP) for scattering $\lambda$ and the localization length $z$. A calculation of the phonon MFP gives an estimate of the distance over which the phonons travel without scattering, and can provide an understanding of the thermal transport process. The line edge roughness scattering limited transmission function $T_{LRS}(\omega)$ is related to the ballistic transmission $T_B(\omega)$, $\lambda(\omega)$, and the channel length $L$ by the relation [48]:

$$T_{\text{LRS}}(\omega) = \frac{\lambda(\omega)}{L+\lambda(\omega)} T_{\text{B}}(\omega). \tag{13}$$

From this, the line edge roughness MFP can be extracted as:

$$\lambda(\omega) = \frac{T_{\text{LRS}}(\omega) L}{T_{\text{B}}(\omega) - T_{\text{LRS}}(\omega)}. \tag{14}$$

When writing down Eq. 13 above, we assume that the channel can be seen as two thermal resistances in series, the channel, and the contacts where the phonons thermalize. Thus, the MFP increases with channel length *L*, until the channel enters the diffusive regime. Strictly speaking, only then does the diffusive MFP converge and can be extracted. While this condition can be reached for short channel lengths for most phonon energies, the acoustic phonons, which carry most of the heat, have very long MFPs, beyond the channel lengths we could simulate. (To provide an indication of the computational cost, we note that a nanoribbon with width of 5nm and a channel of 1 $\mu$m consists of nearly 400,000 atoms. To describe the motion of each atoms a 3×3 matrix is needed, see Eq. 1. The resulting Hamiltonian and Green's functions at each energy point are matrices with a size of 1,200,000×1,200,000. Thus, increasing the length largely increases the computational cost). Therefore, to increase the accuracy in extracting the MFP, we use the transmission values at two different channel lengths as [24]:

$$\lambda(\omega) = \frac{T_{\text{LRS},L_2}(\omega)L_2 - T_{\text{LRS},L_1}(\omega)L_1}{T_{\text{LRS},L1}(\omega) - T_{\text{LRS},L_2}(\omega)}, \tag{15}$$

which accounts partially for the fact that the transmission of phonons with long MFPs has not yet converged fully for the simulated channel length $L$.

In the diffusive regime, the transmission decreases as $1/L$. In the localization regime, on the other hand, for channel lengths greater than the localization length ($z$), the transmission drops exponentially with a characteristic localization length $\zeta$, as [92]:

$$T_{ph}(\omega) \propto \exp\left[-\frac{L}{\zeta(\omega)}\right] \qquad (16)$$

Using a similar reasoning as in the extraction of the diffusive MFP for scattering, we extract the localization length by:

$$\zeta(\omega) = \frac{L_2 - L_1}{ln\left(\frac{T_{LRS,L_1}(\omega)}{T_{LRS,L2}(\omega)}\right)} \qquad (17)$$

where it holds $L_{1,2} \gg \zeta(\omega)$.

Figure 9a shows the average diffusive phonon MFP for scattering on the rough boundaries, $\lambda(\omega)$, as a function of frequency for the channels of two different widths $W$=5nm (red-solid) and $W$=1nm (blue-solid). The MFP is extracted as specified by Eq. 13-15. Since each frequency region, however, enters the diffusive regime at different channel lengths, the MFP for every energy is extracted at the channel length at which the product of the transmission times length ($T \times L$ as in Fig. 3d-f) becomes constant, or levels out. Therefore, Fig. 9 considers a different channel length at all energies for both channel widths, and both $L_1$ and $L_2$ taken at each instance when $T \times L$ levels out. For the wider $W$=5nm channel, the average diffusive MFP (solid-red line) varies from a few 10s of nanometers up to even a few hundreds of nanometers in agreement with Ref. [56] as well. It only drops to a few nanometers around energies $E$~0.16eV due to the large mismatch between the modes in this sparse mode energy region and the formation of a transport gap. For the ultra-narrow $W$=1nm channel (solid-blue line), very large MFPs of the order of several 100s of nanometers are observed for the low frequency phonons close to the zone center originating from the LA modes. This is consistent with the MFP in other carbon nanostructures such as carbon nanotubes and graphene sheets, which is reported to be ~500nm [29, 93, 94, 95], even in the presence of defects [33]. For slightly larger energies, i.e. $E>0.03eV$, the MFP drops sharply to very low values, of at most a few nanometers.

An average MFP value for the entire energy range can be extracted as:

$$\langle\langle\lambda\rangle\rangle = \frac{\int \lambda(\omega)T_{ph}(\omega)W_{ph}(\omega)d\omega}{\int T_{ph}(\omega)W_{ph}(\omega)d\omega} \quad (18)$$

where the phonon window function $W_{ph}(\omega)$ is given by:

$$W_{ph}(\omega) = \frac{3}{\pi^2}\left(\frac{\omega}{k_B T}\right)^2\left(-\frac{\partial n}{\partial \omega}\right) \quad (19)$$

Our calculations show that the average line edge roughness limited diffusive MFP in the case of the narrow GNRs is $\langle\lambda\rangle \sim 30nm$, whereas for the wider GNR of $W$=5nm, it largely increases to $\langle\lambda\rangle \sim 600nm$, also in agreement with other theoretical works [29, 93, 94, 95]. It should be noted that the inclusion of phonon-phonon interaction, which is neglected in this work, can result to smaller MFPs, especially for the high energy optical modes. An accurate modelling of phonon-phonon interaction due to anharmonicities is beyond the scope of this work and will be the subject of our future studies.

In Fig. 9a, we also show the localization length $\zeta(\omega)$ for the narrow $W$=1nm GNR (blue-dashed line). To extract the localization length we use Eq. 17, with $L_1$=500nm and $L_2$=1000nm. The localization length features are very similar to the MFP features. Long localization lengths are observed at very low frequencies, reaching 100s of nanometers. The localization lengths drop to a few nanometers for higher energies. Sharp dips are observed at energies around $E$~0.16eV, which again correlates with the localized features in the $T \times L$ lines of Fig. 3f. In general, $\zeta(\omega)$ and $\lambda(\omega)$ are connected by the Thouless relation $\zeta(\omega)/\lambda(\omega) = N_m$ [96] where $N_m$ is the number of propagating modes in the pristine channel, in our case the same as the value of the ballistic transmission [92].

The ratio $\zeta(\omega)/\lambda(\omega)$ is shown in Fig. 9b for the *W*=1nm GNR (blue-solid line), and as expected, it mostly follows the transmission trend (black-dashed line).

We mention that dephasing mechanisms such as phonon-phonon scattering, could prevent localization, which requires coherence. However, as the localization length is in most of the spectrum smaller than the phonon-phonon scattering MFPs (see Ref. [56]), we expect that localization will be observed in this ultra-narrow channel as described by the drop in *T*×*L* shown in Fig. 3f. Note that we do not attempt to compute the localization lengths for the wider *W*=5nm GNR. This is because from Fig. 3d it is obvious that modes from several parts of the spectrum are not localized at the channel lengths we were able to simulate. However, the large MFPs in this channel suggest even larger localization lengths, in the orders of a few hundreds of nanometers. These lengths are similar to the dephasing lengths, or phonon-phonon scattering MFPs as presented in Ref. [56], and therefore, localization could be prevented. On the other hand, introduction of stronger line-edge roughness amplitude on these wider GNRs would result in smaller roughness scattering MFPs and smaller localization lengths than the ones shown in Fig. 9a (red line). Smaller localization lengths could allow localization to appear, most probably at the same energies as they appear for the *W*=1nm GNR (*E*~0.073V, *E*~0.13eV, and *E*~0.16eV).

The important message to be conveyed from the calculations of $\lambda(\omega)$ and $\zeta(\omega)$, is that phonon transport in ultra-narrow 1D channels consists of multi-scale features, where phonons of MFPs from 100s of nanometers down to a few nanometers are involved. Transport features can vary from ballistic to diffusive and to the localization regimes, depending on the phonon energy, level of disorder, channel length, and channel width. To properly understand phonon transport in 1D channels all of these features need to be taken into proper consideration.

## VI. Thermal conductivity

Finally, it is important to extend the analysis in including features of thermal conductivity in ultra-narrow GNRs. The thermal conductivity of the GNR channels is a length dependent quantity and calculated using the thermal conductance as $\kappa_l = LK_l / A$, where *A* is the cross sectional area of the GNR with its height assumed to be 0.335nm. Figure 10 shows the thermal conductivity versus channel length for GNRs with width *W*=5nm (red-diamond line) down to *W*=1nm (blue-circle line). The increase in thermal conductivity with channel length for short channels, and saturation for the longer ones, indicates the transition between ballistic and diffusive transport which was also observed at various instances [56]. For the wider GNR channels, the saturation begins for length scales of several hundreds of nanometers. At this channel length, however, the narrower GNR with *W*=1nm is already driven into the localization regime (blue line). Ballistic transport dictates that the thermal conductivity increases linearly with channel length, while saturation comes due to scattering. The strength of the line edge roughness is indicated by the deviation from unity of the slope of the thermal conductivity lines for short channel lengths [97, 98]. A power law behavior $L^\alpha$ is expected for 1D channels [97, 98]. From our calculations, for the wider channels *W*=4nm and 5nm the slope is *α*=0.7. As the width decreases, the slope decreases as well, with the *W*=3nm having *α*=0.65, and the narrowest channel *W*=1nm having *α*=0.5.

## VI. Conclusions

In this work we have investigated the thermal transport properties of low-dimensional, ultra-narrow graphene nanoribbon (GNR) channels under the influence of line edge roughness disorder. We employed the non-equilibrium Green's function (NEGF) method for phonon transport and the force constant method for the description of the phonon modes. We show that the effect of line edge roughness affects different parts of the spectrum in different ways: i) Under strong effective disorder, the thermal conductivity is dominated by the low frequency acoustic modes, which have MFPs of several hundred nanometers and suffer from localization only under extreme confinement in purely 1D channels. At ultra-narrow channel widths they tend to completely dominate thermal transport; ii) Regions of the spectrum with a dense population of modes such as the optical modes, can contribute significantly to thermal transport, even if their group velocity is low; iii) Regions of the spectrum with low mode density end up becoming effective transport gaps as the length of the channel increases, or the width decreases, and contribute little to thermal transport, even if they are relatively dispersive; iv) Regions of the spectrum with very low mode densities, populated with relatively 'flat' modes suffer from band mismatch in the presence of both confinement or roughness, which creates even stronger transport gaps and completely eliminate their ability to carry heat. In general, confinement reduces the population of the modes in the entire energy spectrum (except the low frequency acoustic regions), and under the influence of disorder they fall into category (iv), i.e. confinement and roughness reduces phonon transmission by introducing effective transport gaps and band mismatch. This drives transport at those energies into the localization regime. Finally, we show that although the transmission of several energy regions is severely degraded in the presence of line edge roughness, for channels with lengths up to $L$=1μm that we have simulated, only the overall thermal conductivity of the ultra-narrow $W$=1nm GNRs is driven into the localization regime.

# Acknowledgements

Mahdi Pourfath and Hans Kosina were supported by the Austrian Science Fund (FWF) contract P25368-N30. Hossein Haramitaheri and Neophytos Neophytou acknowledge Dr Mischa Thesberg for useful discussions. The computational results presented have been achieved in part using the Vienna Scientific Cluster (VSC).

Appendix:

In the entire paper we use armchair GNRs (AGNRs). Here we plot the corresponding corresponding phonon dispersion (Fig. A1), and transmission probability (Fig. A2), for zig-zag edge GNRs (ZGNRs). These are the corresponding Fig. 1, and Fig. 6a for AGNRs in the manuscript. In both figures, the results for ZGNRs are very similar to those for AGNRs. Strong reductions in the transmission function around $E$=0.07eV, $E$=0.11eV and $E$=0.16eV are observed (Fig. A2). In the case of ZGNRs, however, the transmission around $E$=0.07eV and $E$=0.11eV is reduced much less compared to AGNRs (see Fig. 6a). This is attributed to the slight differences in the phonon dispersion relations of AGNR versus ZGNR, observed if one compares Fig. A1 with Fig. 1. As the GNR width is reduced from $W$=5nm to $W$=1nm, the 'empty regions' in the dispersion of the ZGNR (or the 'effective bandgap' formation regions), are not as distinctive as in the case of the AGNRs analyzed in the paper. ZGNRs have slightly more dispersive bands, something also validated by first principle calculations [20], which: i) make the ballistic thermal conductance of a ZGNR higher than that of its AGNR counterpart (ZGNR transmission is in general higher than the AGNR transmission), and ii) does not allow the formation of 'effective bandgaps' upon confinement and roughness as easily.

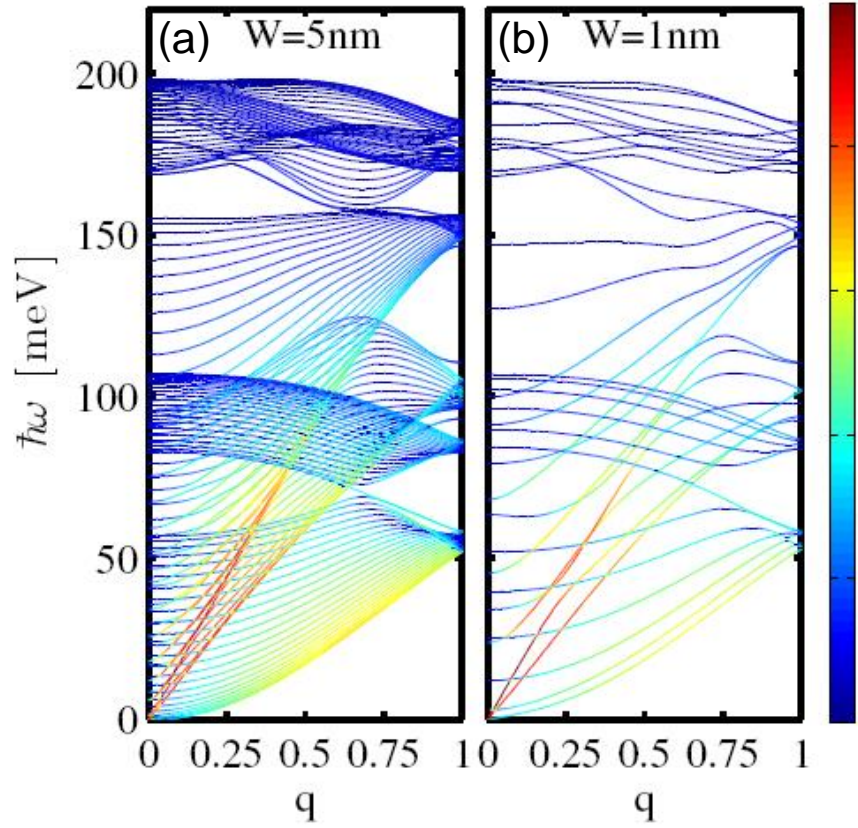

Figure A1: Phonon dispersions for (a) *W*=5nm, (b) *W*=1nm wide zig-zag nanoribbons (ZGNRs). As the width is decreased, the number of phonon modes is also reduced. The colormap shows the contribution of each phonon state to the total ballistic thermal conductance (red: largest contribution, blue: smallest contribution). (This is the corresponding ZGNR case as Fig. 1 in the manuscript is for armchair ribbons - AGNRs)

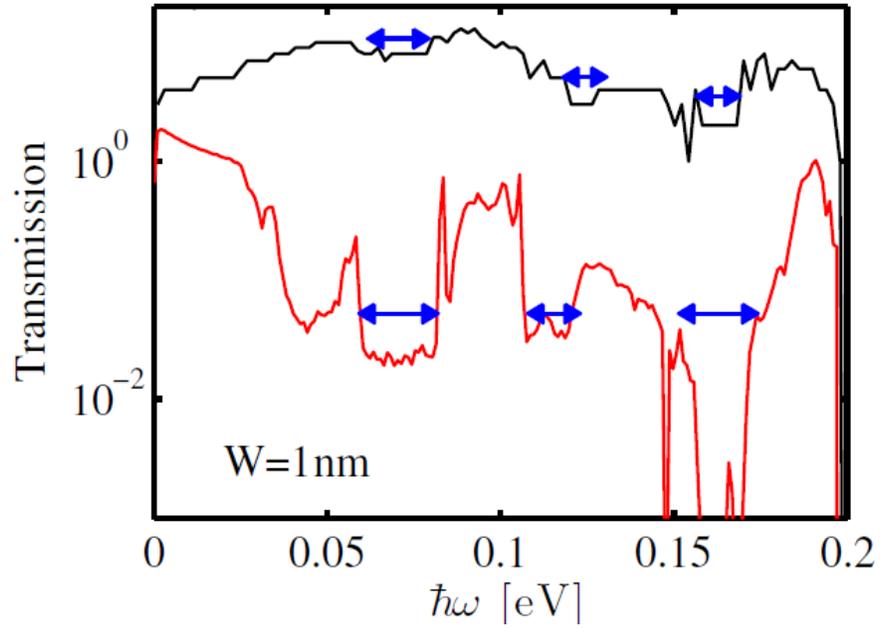

Figure A2: The transmission function versus energy in logarithmic scale for rough edge zig-zag GNRs of width $W$=1nm. The ballistic transmission (pristine GNRs, non-roughened ribbons) is depicted by the black line. Nanoribbons with length $L$=40nm is shown by the red line. (This is the corresponding ZGNR case as Fig. 6a in the manuscript is for armchair ribbons - AGNRs)

Figure 1:

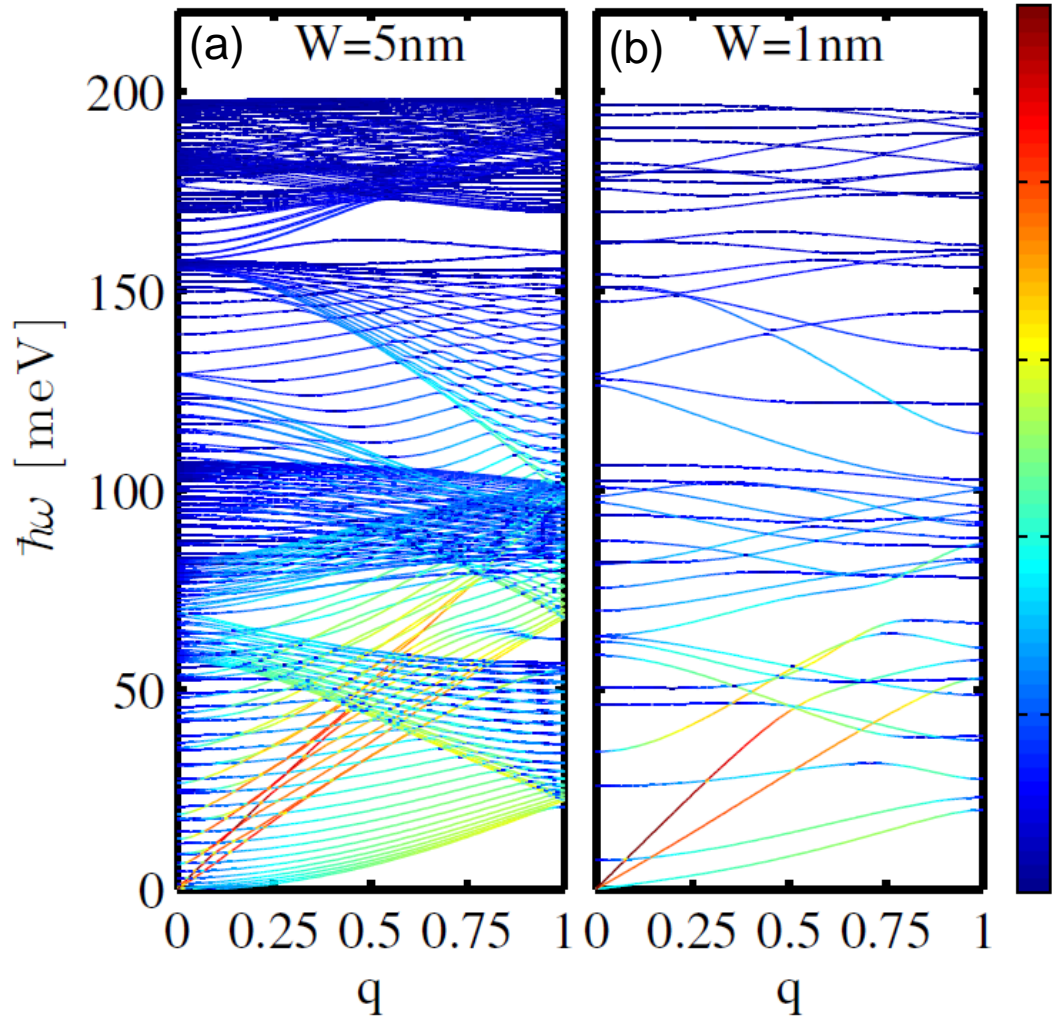

Figure 1 caption:

Phonon dispersions for (a) $W$=5nm, (b) $W$=1nm wide armchair nanoribbons. As the width is decreased, the number of phonon modes is also reduced. The colormap shows the contribution of each phonon state to the total ballistic thermal conductance (red: largest contribution, blue: smallest contribution).

Figure 2:

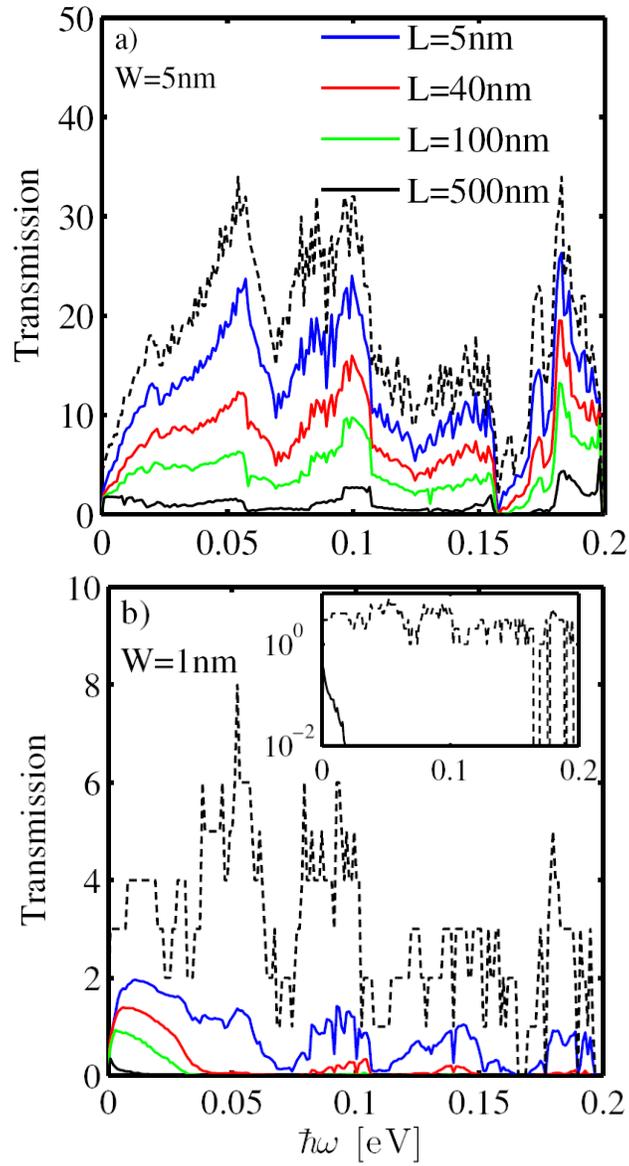

## Figure 2 caption:

The transmission function versus energy for rough edge GNRs of width (a) *W*=5nm and (b) *W*=1nm. Nanoribbon lengths of *L*=5nm (blue lines), *L*=40nm (red lines), *L*=100nm (green lines), and *L*=500nm (black lines) are considered. The ballistic transmissions (pristine, non-roughened ribbons) are depicted in black-dashed lines.

Figure 3:

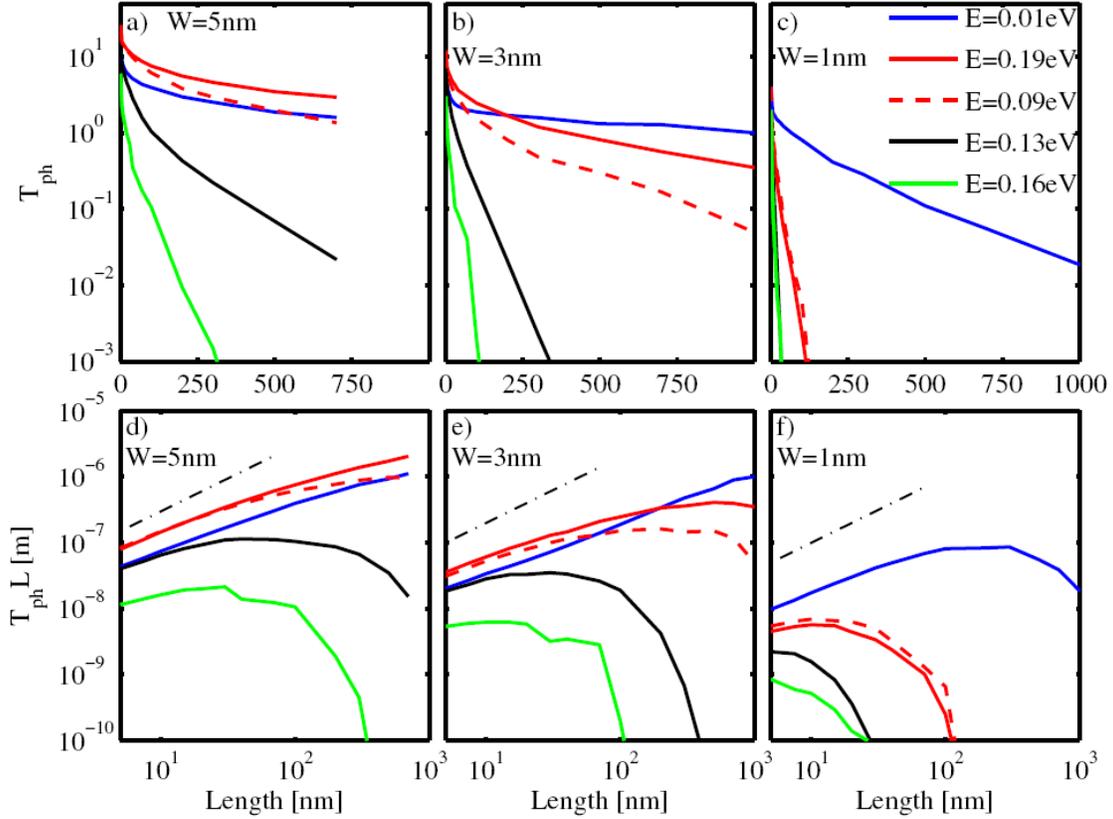

# Figure 3 caption:

(a-c) The phonon transmission of rough nanoribbons of widths $W$=5nm (a), $W$=3nm (b) and $W$=1nm (c) for specific energies versus channel length. Energies $E$=0.01eV (blue lines) correspond to the acoustic branches. $E$=0.19eV and $E$=0.09eV (red-solid and red-dashed lines respectively) correspond to regions of the spectrum where the bands are numerous, but mostly 'flat'. $E$=0.16eV (green line) corresponds to a region of the spectrum at the interface between dispersive and flat bands, in which narrow bandgaps are formed as the width is reduced. $E$=0.13eV (black line) corresponds to a spectrum region where dispersive bands exist, but as the width is reduced they are reduced in number and in addition narrow bandgaps form. (d-e) The phonon transmission times the channel length $T{\times}L$ for the same situations as in (a-c).

Figure 4:

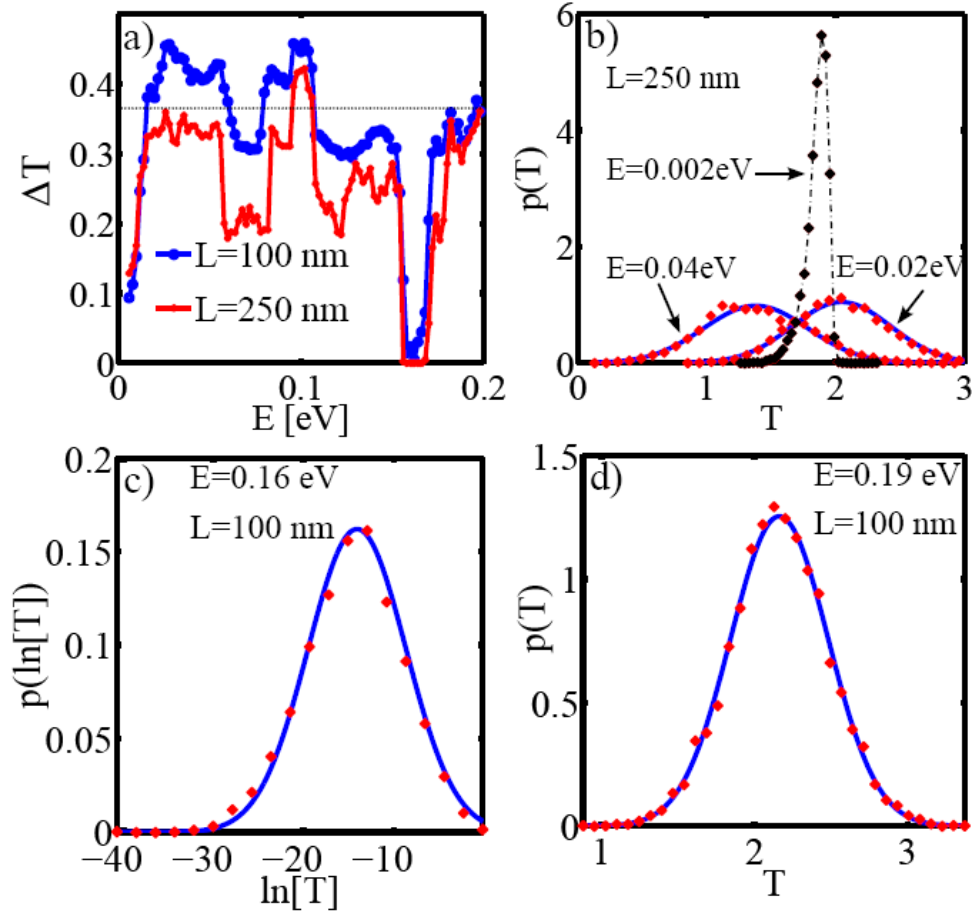

Figure 4 caption:

(a) The fluctuation of phonon transmission as a function of energy for GNRs with a width of $W$=3 nm and lengths of $L$=100 nm (blue) and $L$=250 nm (red). (b) The histogram of phonon transmission at $E$=0.002eV (black dots), $E$=0.02eV and $E$=0.04eV for GNRs of lengths $L$=250 nm. (c) The histogram of phonon transmission at $E$=0.16 eV for GNR lengths $L$=100 nm. The distribution shown in logarithmic scale follows a Gaussian distribution, which is equivalently a log-normal distribution function. (d) The histogram of phonon transmission at $E$=0.19 eV for GNRs with lengths $L$=100 nm. The histogram follows a Gaussian distribution function, a characteristic of diffusive transport regime. The blue lines in (b), (c), and (d) are Gaussian fitted lines using the average and standard deviation of the simulation results.

Figure 5:

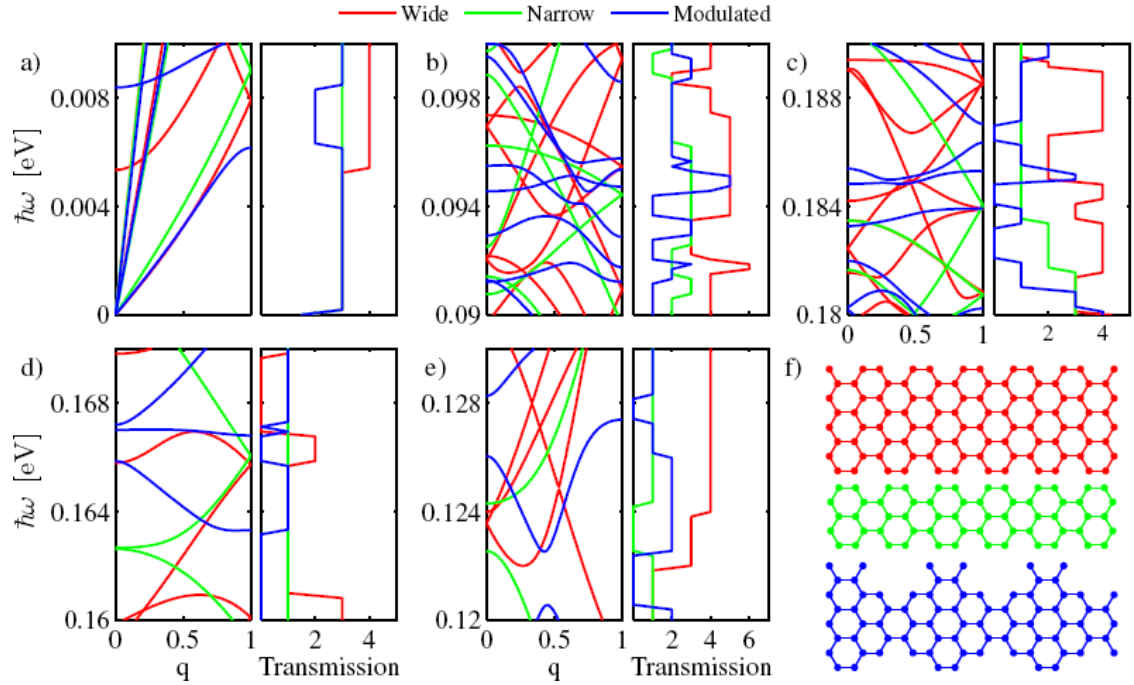

## Figure 5 caption:

The phonon dispersion and transmission function of the $W$=1nm GNR under three different situations as shown in sub-figure (f). i) A slightly wider channel of $W$=1.11nm (red), ii) a slightly narrower channel of $W$=0.74nm (green), and iii) a GNR whose width is periodically modulated (blue) are considered. The latter mimics a rough ribbon. Different sets of energies are shown: (a) $E$=0eV to $E$=0.01eV (acoustic modes). (b) $E$=0.09eV to $E$=0.1eV (optical modes). (c) $E$=0.18eV to $E$=0.19eV (optical modes). (d) $E$=0.16eV to $E$=0.17eV (regions between 'quasi-acoustic' and optical modes). (e) $E$=0.12eV to $E$=0.13eV ('quasi-acoustic' modes). (f) Schematic of the atomistic geometries of the three nanoribbon cases.

Figure 6:

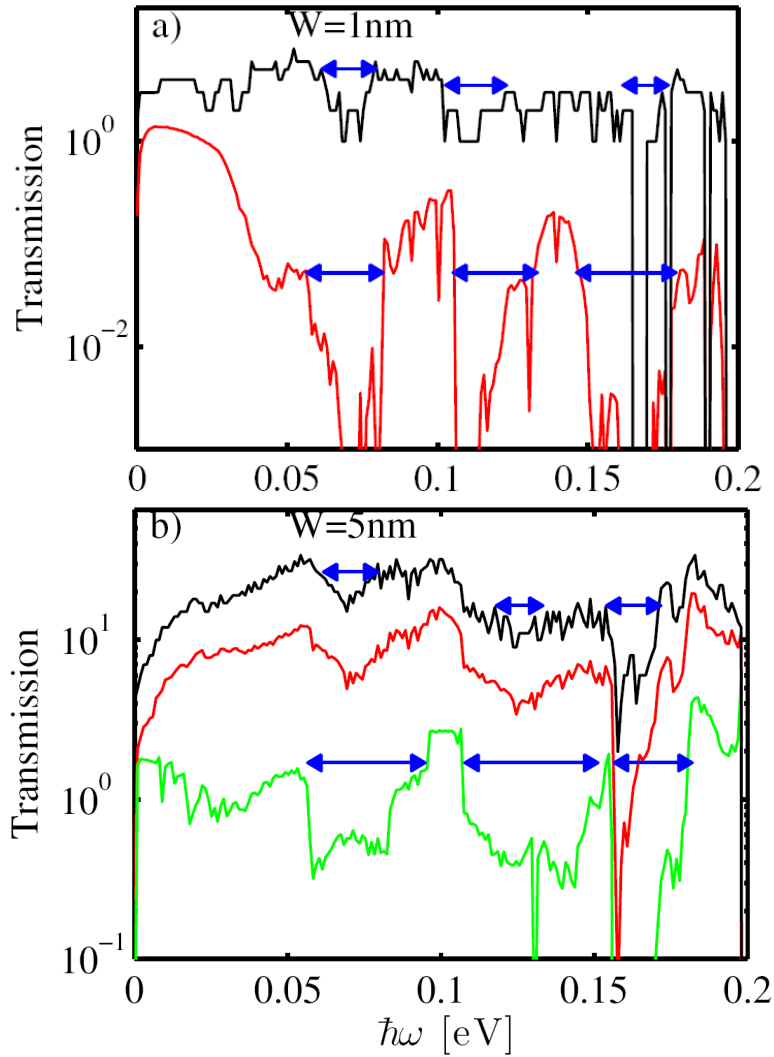

## Figure 6 caption:

The transmission function versus energy in logarithmic scale for rough edge GNRs of width (a) $W$=1nm and (b) $W$=5nm. The ballistic transmission (pristine GNRs, non-roughned ribbons) is depicted by the black lines. The transmission of nanoribbons with length $L$=40nm is shown by the red lines. In (b) the transmission of the GNR with length $L$=500nm is also shown in green.

Figure 7:

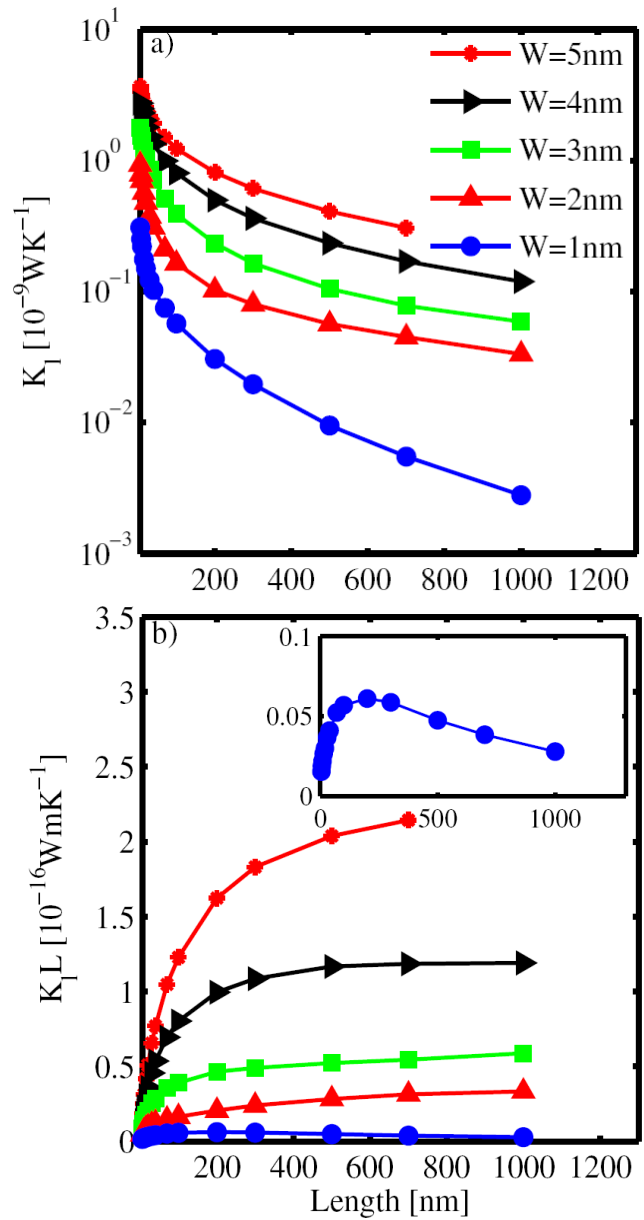

## Figure 7 caption:

(a) The thermal conductance versus channel length of rough GNRs with widths $W$=5nm (red-crosses), $W$=4nm (black-triangles), $W$=3nm (green-squares), $W$=2nm (red-triangles), and $W$=1nm (blue-circles) are shown. (b) The same channels as in (a), but the thermal conductance times the channel length $K_l \times L$ is shown. Inset of (b): Zoom-in for the $W$=1nm case.

Figure 8:

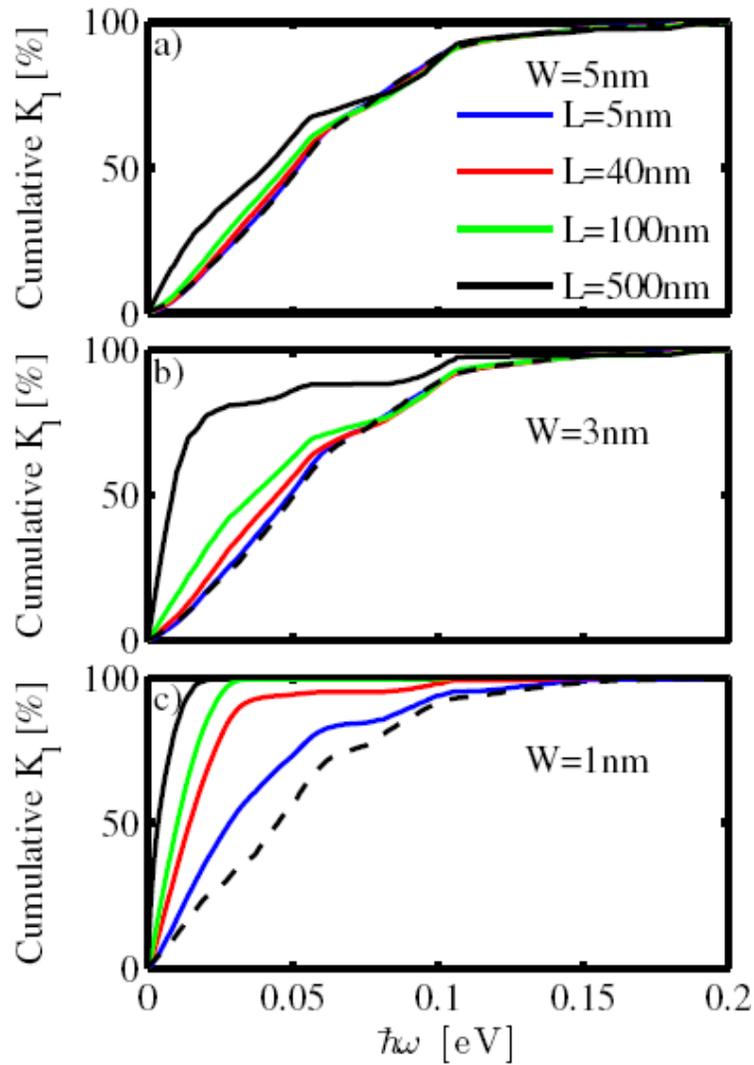

Figure 8 caption:

The cumulative thermal conductance versus energy for GNR channels of different widths. (a) $W$=5nm, (b) $W$=3nm, and (c) $W$=1nm. For every case, the dashed line indicates the ballistic case. Channel lengths of $L$=5nm (blue), $L$=40nm (red), $L$=100nm (green), and $L$=500nm (black) are shown.

Figure 9:

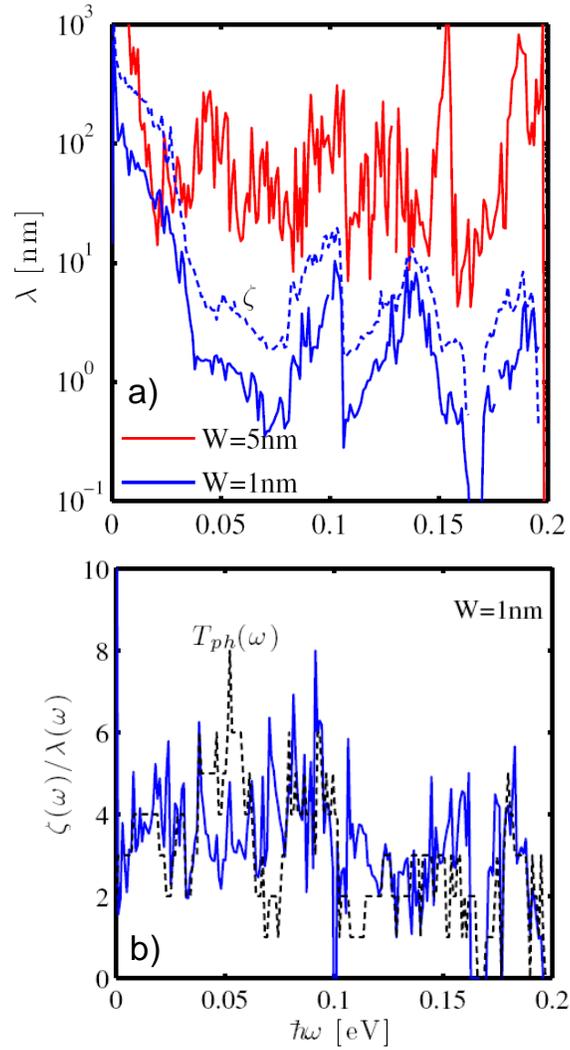

Figure 9 caption:

(a) The average diffusive transport mean-free-path (solid) versus energy of rough GNRs of widths $W$=5nm (red) and $W$=1nm (blue). The MFP as a function of energy is extracted at the channel length at which the $T \times L$ product is constant; therefore, the channel length differs for each energy. The dashed-blue line shows the localization length for the $W$=1nm. (b) The ratio of the localization length $\zeta(\omega)$ over the MFP $\lambda(\omega)$ for the $W$=1nm rough ribbon of length $L$=1000nm (blue line) and the transmission probability of the pristine $W$=1nm GNR (black-dashed line).

Figure 10:

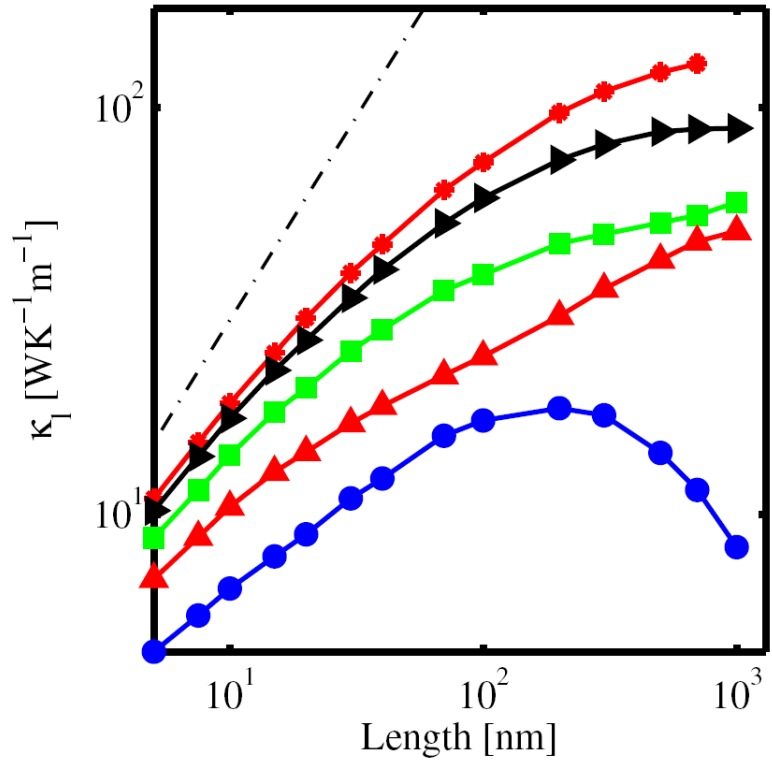

## Figure 10 caption:

The thermal conductivity versus channel length for GNRs with widths $W$=5nm (red-crosses), $W$=4nm (black-triangles), $W$=3nm (green-squares), $W$=2nm (red-triangles), and $W$=1nm (blue-circles).